\documentclass{article}

\PassOptionsToPackage{numbers, compress}{natbib}

\usepackage[final]{neurips_2025}

\usepackage[utf8]{inputenc} 
\usepackage[T1]{fontenc}    
\usepackage{hyperref}       
\usepackage{url}            
\usepackage{booktabs}       
\usepackage{amsfonts}       
\usepackage{pifont}         
\usepackage{nicefrac}       
\usepackage{microtype}      
\usepackage{xcolor}         
\usepackage{multirow}

\usepackage{graphicx}
\usepackage{amsmath}
\usepackage{pifont}
\usepackage{bm}

\usepackage{arydshln}
\usepackage{authblk}

\title{CLEAR: Continuous Latent Autoregressive Modeling for High-quality and Low-latency Speech Synthesis}

\author[$^*$,1]{\textbf{Chun Yat Wu}} 
\author[$^*$,2]{\textbf{Jiajun Deng}} 
\author[1]{\textbf{Guinan Li}} 
\author[1]{\textbf{Qiuqiang Kong}} 
\author[2]{\textbf{Simon Lui}} 

\affil[1]{The Chinese University of Hong Kong}
\affil[2]{Hong Kong Research Center, Huawei}

\begin{document}
\maketitle

\renewcommand{\thefootnote}{\fnsymbol{footnote}} 
\setcounter{footnote}{0} 
\footnotetext{* These two authors contributed equally to this work.} 
\vspace{-0.5cm}
\begin{abstract}
Autoregressive (AR) language models have emerged as powerful solutions for zero-shot text-to-speech (TTS) synthesis, capable of generating natural speech from a few seconds of audio prompts. However, conventional AR-based TTS systems relying on discrete audio tokens
face the challenge of lossy compression during tokenization, requiring longer discrete token sequences to capture the same information as continuous ones, which adds inference latency and complicates AR modeling. To address this challenge, this paper proposes the \textbf{C}ontinuous \textbf{L}at\textbf{e}nt \textbf{A}utoreg\textbf{r}essive model (CLEAR), a unified zero-shot TTS framework that directly models continuous audio representations. More specifically, CLEAR introduces an enhanced variational autoencoder with shortcut connections, which achieves a high compression ratio to map waveforms into compact continuous latents. A lightweight MLP-based rectified flow head that operates independently for each hidden state is presented to model the continuous latent probability distribution, and trained jointly with the AR model within a single-stage framework. 
Experiments show that the proposed zero-shot CLEAR TTS can synthesize high-quality speech with low latency. Compared to state-of-the-art (SOTA) TTS models, CLEAR delivers competitive performance in robustness, speaker similarity and naturalness, while offering a lower real-time factor (RTF). In particular, CLEAR achieves SOTA results on the LibriSpeech test-clean dataset, with a word error rate of 1.88\% and an RTF of 0.29. Moreover, CLEAR facilitates streaming speech synthesis with a first-frame delay of 96ms, while maintaining high-quality speech synthesis. Audio samples are available at \href{https://anonymous.4open.science/w/clear_demo_/}{here}.
\end{abstract}

\vspace{-0.5cm}
\section{Introduction}
Recent advances in autoregressive (AR) language models (LMs) have demonstrated impressive generative capabilities in the field of speech synthesis \cite{borsos2023audiolm,chen2024valle2neuralcodec,wang2023neural,wang2025spark,ye2025llasa,zeghidour2021soundstream}, driven by their ability to effectively model sequences and leverage scaling properties. With a few seconds of audio as a prompt, current AR-based text-to-speech (TTS) systems that consecutively predict the next discrete token based on previous tokens as condition can synthesize speech for any input text while accurately mimicking the speaker's characteristics from a few seconds of a audio prompt \cite{du2024cosyvoicescalablemultilingualzeroshot,du2024cosyvoice2scalablestreaming,jia2025ditardiffusiontransformerautoregressive,meng2024autoregressivespeechsynthesisvector}. 

Despite their promising zero-shot TTS capability, AR modeling of discrete tokens faces two fundamental limitations in delivering high-quality and low-latency speech synthesis. \textbf{First}, lossy compression during audio tokenization restricts reconstruction quality, as current neural audio codecs often require high bitrates to achieve fidelity comparable to continuous representations \cite{blau2019rethinking,defossezHighFidelityNeural2022}. For instance, encoding one second of high-fidelity audio at 16 kHz typically requires hundreds of audio tokens with a codebook of size 1024, resulting in long sequences that not only complicate AR modeling but also lead to high inference latency \cite{wu2024towards}. To fill the aforementioned information gap, most state-of-the-art TTS systems, such as CosyVoice \cite{du2024cosyvoicescalablemultilingualzeroshot} and Seed-TTS \cite{anastassiou2024seed}, often adopt a two-stage cascade of generative models: a LM initially generates discrete audio tokens, which are subsequently refined using token-based diffusion models. However, this cascaded design is limited by error accumulation, misaligned optimization objectives, and substantial computational demands. \textbf{Second}, neural audio codecs are challenging to train and highly sensitive to gradient approximation techniques in discrete distributions. For example, VQ-GANs \cite{esserTamingTransformersHighResolution2021} require specialized techniques, such as auxiliary losses and codebook re-initialization \cite{huh2023straightening}, to achieve effective training.

An intuitive method to overcome these limitations is to model directly in continuous audio representations, thereby eliminating the need for discrete-valued tokenizers. However, two essential factors need to be addressed to achieve a high-quality, low-latency TTS system.
\begin{itemize}
    \item \textbf{Probability Distribution Modeling:} Compact continuous representations carry rich and complex information over discrete tokens, requiring autoregressive models with advanced capabilities to handle them effectively. For example, the regression-based loss functions used in MELLE \cite{meng2024autoregressivespeechsynthesisvector} (i.e., mean squared error), which are based on simplified distributional assumptions, often struggle to capture complex speech patterns, leading to blurry and overly uniform predictions \cite{ren2022revisiting}.

    \item \textbf{Training/Inference Efficiency:} While diffusion techniques have proven effective in modeling continuous representations \cite{chen2024f5ttsfairytalerfakesfluent} and can deliver impressive performance when integrated with AR models \cite{li2024autoregressiveimagegenerationvector}, it comes at the cost of reduced training and inference efficiency. This limitation hampers the low-latency requirements of TTS systems. For example, ARDiT \cite{liu2024autoregressivediffusiontransformertexttospeech} introduces considerable computational overhead and high inference latency by forcing adjustments to the language model parameters to incorporate the diffusion process.
\end{itemize}
To this end, this paper presents a \textbf{C}ontinuous \textbf{L}at\textbf{e}nt \textbf{A}utoreg\textbf{r}essive model (CLEAR) with rectified flow for high-quality and low-latency text-to-speech synthesis. CLEAR is a robust zero-shot TTS model that operates in a single pass, predicting continuous autoencoder (AE) latents autoregressively conditioned on prior latents and text tokens. In response to the challenges associated with AR modeling of continuous audio representations: \textbf{1)} A diffusion network is utilized to model the underlying continuous latent probability distribution, conditioned on the hidden state produced by the autoregressive model. Its ability to estimate arbitrary densities ensures the preservation of natural speech's complex characteristics. The diffusion network and autoregressive model are jointly trained under a single-stage framework using the rectified flow loss criterion. \textbf{2)} To facilitate the training and inference efficiency, an enhanced variational auto-encoder (VAE) is utilized to map the audio waveform to the continuous audio latent. Specifically, the enhanced VAE not only delivers high-quality audio reconstruction but also achieves significantly higher compression rates than traditional mel-spectrogram feature extraction in continuous representation modeling, substantially reducing computational overhead, storage requirements, and enhancing inference efficiency. Moreover, compact VAE latents exhibit greater training stability and require less computation resources than mel spectrograms in AR modeling. \textbf{3)} To further minimize first-frame latency during inference, the diffusion module adopts a simple MLP architecture and and a causal VAE decoder. In contrast to complex Diffusion Transformer (DiT) models, which must wait for the full generation of autoregressive hidden states before starting denoising, the MLP-based diffusion module allows the diffusion process to operate independently for each hidden state and immediately upon receiving the first hidden state, facilitating real-time streaming speech synthesis. The main contributions of this paper are summarized as follows:
\begin{itemize}
    \item This paper presents CLEAR, a novel zero-shot TTS framework that directly models continuous audio representations, overcoming the limitations of lossy compression and high inference latency in discrete-valued AR-based TTS systems \cite{chen2024valle2neuralcodec,du2024cosyvoicescalablemultilingualzeroshot,guo2024fireredtts,han2024vall,lajszczakBASETTSLessons2024,nishimura2024hall,song2025ella,xin2024rall,wang2023neural,zhang2023speak}.
    
    \item This paper introduces an enhanced VAE that transforms audio waveforms into compact continuous audio representations, which offers the dual benefits of improved training efficiency and reduced inference latency. In contrast, previous approaches rely on lengthy mel-spectrogram feature sequences \cite{meng2024autoregressivespeechsynthesisvector,wang2025felleautoregressivespeechsynthesis}, which are computationally inefficient and increase the complexity of AR modeling.
    
    \item  This paper presents a simplified rectified flow head for continuous latent probability distribution modeling, jointly trained with an autoregressive model in a single-stage framework. In contrast, prior continuous-valued AR-based TTS systems either rely on simple distributional assumptions \cite{meng2024autoregressivespeechsynthesisvector}, which fail to effectively model complex speech patterns, or adopt complex DiT networks \cite{jia2025ditardiffusiontransformerautoregressive,liu2024autoregressivediffusiontransformertexttospeech} that incur significant inference latency, hindering their suitability for real-time streaming TTS.

    \item Experimental results demonstrate that the proposed zero-shot CLEAR TTS model can synthesize high-quality speech with low latency. It achieves competitive performance in robustness, speaker similarity, and naturalness, while offering fast inference compared to state-of-the-art TTS models. Moreover, CLEAR provides streaming speech synthesis with a first-frame speech synthesis delay as low as 96ms, without compromising speech quality.
\end{itemize}

\section{Related Work}
\textbf{Zero-shot TTS.}
In contrast to traditional speech synthesis methods that require hours of high-quality transcribed data from the target speaker, zero-shot TTS synthesizes the voice of a new speaker with a few seconds of a voice prompt. Existing zero-shot TTS systems can be spearheaded into two paradigms based on their output generation mechanisms: autoregressive and non-autoregressive (NAR) paradigms. Autoregressive LM represented by VALL-E \cite{wang2023neural}, predict audio representations sequentially, conditioned on text tokens and previously generated audio representations. In contrast, NAR models represented by Voicebox generate the entire audio sequence representations in parallel \cite{le2023voicebox}. Diffusion models, in particular, have played a crucial role in the success of NAR TTS systems, including but not limited to F5-TTS \cite{chen2024f5ttsfairytalerfakesfluent}, Naturalspeech \cite{juNaturalSpeech3ZeroShot2024,shenNaturalSpeech2Latent2023,tanNaturalSpeechEndtoEndTexttoSpeech2024}, E2-TTS \cite{eskimez2024e2ttsembarrassinglyeasy}, and E3-TTS \cite{gaoE3TTSEasy2023}.

\textbf{Discrete-valued AR-based TTS.}
Autoregressive language models (LMs) have exhibited significant zero-shot and in-context learning capabilities in text-to-speech synthesis, where discrete tokenization is commonly utilized. VALL-E \cite{wang2023neural} pioneered the approach of treating TTS as a conditional language modeling task by compressing audio waveforms into discrete tokens using a neural audio codec encoder, predicting these tokens autoregressively, and reconstructing the waveforms through a codec decoder. The success of VALL-E's zero-shot capabilities has sparked widespread interest in discrete neural codec language models, targeting improvements in multilingual generalization \cite{zhang2023speak}, inference efficiency \cite{chen2024valle2neuralcodec}, and advancing performance \cite{han2024vall,nishimura2024hall,song2025ella,xin2024rall}. Despite this, bitrate limitations during audio tokenization often prevent discrete representations from reconstructing complex speech patterns with high fidelity. To this end, models such as CosyVoice \cite{du2024cosyvoicescalablemultilingualzeroshot, du2024cosyvoice2scalablestreaming}, BASE-TTS \cite{lajszczakBASETTSLessons2024}, FireRed TTS \cite{guo2024fireredtts}, and Seed-TTS \cite{anastassiou2024seed} employ a two-stage training framework. In the first stage, a LLM maps text to coarse-grained discrete tokens using cross-entropy loss. In the second stage, a conditional flow-matching model refines these tokens into fine-grained mel-spectrograms using diffusion loss, resulting in improved synthesized speech quality. Nonetheless, the inherent limitations of two-stage training require extra efforts addressing issues such as inference latency and error accumulation. 

\textbf{Continuous-valued AR-based TTS.}
Recent advances in continuous representation-based TTS systems have shown promising performance by eliminating the need for intricate codec training. For instance, MELLE \cite{meng2024autoregressivespeechsynthesisvector} autoregressively predicts Mel spectrograms parameterized by a Gaussian distribution, but its simplified distribution assumptions limit its ability to model complex speech patterns. To address such limitations, several studies have explored diffusion techniques to model multimodal distributions in continuous representations. ARDiT \cite{liu2024autoregressivediffusiontransformertexttospeech} leverages a text-conditioned DiT with a special masking mechanism to sample multiple tokens, but regrading Transformer as denosing backbone incurs significant computational overhead. DiTAR \cite{jia2025ditardiffusiontransformerautoregressive} mitigates this by aggregating speech tokens into patches for input to a language model and using a LocDiT module for reconstruction. However, its reliance on transformers as the diffusion backbone, due to the attention mechanism, limits its ability to produce streaming speech output efficiently. Similarly, FELLE \cite{wang2025felleautoregressivespeechsynthesis} incorporates information from the previous step to refine the general prior distribution in flow matching, while SALAD \cite{turetzkyContinuousSpeechSynthesis2024} enhances modeling fidelity by employing latent diffusion model per-token  on continuous representations. Despite these advancements, the two-stage coarse-to-fine generation process in FELLE and the additional text-to-semantic module in SALAD for contextual guidance and stopping control not only lead to cumbersome training framework but also increase computational demands. 

\begin{figure}[htbp]
    \centering
    \includegraphics[width=1.0\textwidth]{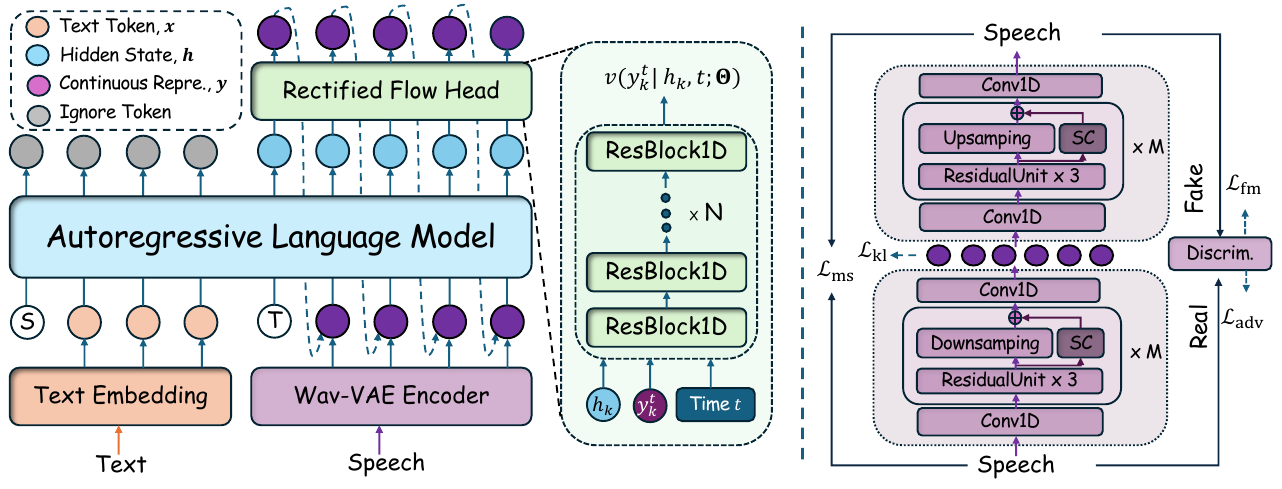}
    \caption{Overview of the CLEAR architecture (left) and the enhanced wav-VAE architecture (right). The CLEAR architecture comprises the wav-VAE encoder (purple), the autoregressive (AR) model (blue), and the rectified flow head (green). Dashed lines indicate AR decoding during inference. "S", and "T" represent the "start of sequence" and "turn of speech" respectively. The enhanced wav-VAE architecture incorporates shortcut connections (SC) in both the downsampling and upsampling layers of the VAE encoder and decoder.}
    \vspace{-0.3cm}
    \label{fig:label1}
\end{figure}
\section{Method}
\subsection{Continuous-valued Autoregressive Modeling for Text-to-Speech Synthesis}
CLEAR formulates text-to-speech synthesis as an autoregressive language modeling task through next-token prediction. Given a sequence of continuous audio representations $\bm{y} = (\bm{y}_1, \bm{y}_2,\cdots,\bm{y}_K)$ that $\bm{y}_i\in\mathbb{R}^{d}$ and its corresponding sequence of text tokens $\bm{x}=(x_1,x_2,\cdots,x_N)$, where $K$ and $N$ are the number of continuous audio tokens and text tokens respectively, the model is expected to predict the joint distribution of the continuous audio sequence conditioned on text tokens, which can be factorized into 
\begin{equation}
p(\bm{y}|\bm{x};\bm{\Theta}) = \prod_{k=1}^{K} p(\bm{y}_k \mid \bm{y}_{<k}, \bm{x}; \bm{\Theta})\label{eq1},
\end{equation}
where $\bm{y}_{<k}=(\bm{y}_1,\bm{y}_2,\cdots,\bm{y}_{k-1})$ denotes the audio representation sequence before the current autoregressive step $k$, and $\bm{\Theta}$ denotes the model parameters. As shown in Figure~\ref{fig:label1}, in the CLEAR framework, the generation can be separated into two parts ${\bm \Theta}=\{{\bm \Theta}_{\text {LM}}, {\bm \Theta}_{\text {RF}} \}$, where ${\bm \Theta}_{\text {LM}}$ denotes the language model parameters responsible for predicting the conditioning vector $\bm{h}_k=f(\bm{y}_{<k},\bm{x};{\bm \Theta}_{\text {LM}})$, ${\bm \Theta}_{\text {RF}}$ denotes the MLP rectified flow head parameters that perform the probability of the next token by $p(\bm{y}_{k}|\bm{h}_t;{\bm \Theta}_{\text {RF}})$. Then the joint distribution in Eqn.~(\ref{eq1}) can be reformulated as
\begin{equation}
p(\bm{y}|\bm{x};\bm{\Theta}) = \prod_{k=1}^{K} p(\bm{y}_k \mid \bm{h}_{k};{\bm \Theta}_{\text {RF}})=\prod_{k=1}^{K} p(\bm{y}_k \mid f(\bm{y}_{<k},\bm{x};{\bm \Theta}_{\text {LM}});{\bm \Theta}_{\text {RF}}),
\end{equation}
In this paper, the language model $f(\cdot;{\bm \Theta}_{\text{LM}})$ is built on a unidirectional Transformer decoder to capture the dependency between semantic and acoustic information. All Transformer layers adopt the Pre-Norm configuration \cite{xiong2020layernormalizationtransformerarchitecture} and integrate RMSNorm \cite{zhang2019rootmeansquarelayer}, RoPE \cite{su2023roformerenhancedtransformerrotary} and SwiGLU \cite{shazeer2020gluvariantsimprovetransformer} operations to align with current LLM designs. The LM output $\bm{h}_k$, serving as a conditioning vector, is passed to the diffusion module to synthesize the next-frame continuous representation ${\bm y}_k$. Note that the language model is jointly trained with the rectified flow head within a single framework. Further details are provided in the following section.

\subsection{Probability Distribution Modeling}
The denoising diffusion model, capable of representing arbitrary probability distributions, is utilized to model the distribution of each token $p(\bm{y}_{k}|\bm{h}_k;{\bm \Theta}_{\text {RF}})$. The core idea behind diffusion operation is to model the data distribution by reversing a forward noising process. Unlike traditional diffusion operation model the noise estimation score function at various noise levels, rectified flow (RF) \cite{liu2022flowstraightfastlearning} provides a more direct framework by learning a smooth trajectory that rectifies the corrupted data distributions into the true data distribution. Specifically, it defines the forward process as straight-line paths between the data distribution and a standard normal distribution. At the $k$-th autoregressive step, the data point at the diffusion time step $t$ can be expressed as ${\bm y}_k^{t}=(1-t){\bm y}_k^{0}+t{\bm y}_k^{1}$, where ${\bm y}_k^1=\bm{y}_k$ and ${\bm y}_k^{0}\sim{\cal N}(\bm{0},\bf {I})$ is a noise vector drawn from the normal distribution. 
The rectified flow, characterized by an ordinary differential equation (ODE), is defined as $d \bm{y}_k^{t} = v(\bm{y}_k^{t}, t){d t}$, where $v(\bm{y}_k^{t}, t)$ represents the drift vector field. The loss function for the underlying probability distribution $p(\bm{y}_{k}|\bm{h}_k;{\bm \Theta}_{\text {RF}})$ can then be expressed as a denoising criterion, formulated as
\begin{equation}
{\cal L}_{\text {RF}}(\bm y_k,\bm h_k) = \mathbb{E}_{t,{\bm y}_k^{0}} \left[ \lVert {\bm u}_k - v(\bm{y}_k^{t}|{\bm h}_k,t; \bm\Theta) \rVert^2 \right],
\end{equation}
where ${\bm u}_k=({\bm y}_k^{1}-{\bm y}_k^0)$ denotes the true conditional vector field for the original data distribution in continuous audio representation. The notation $v(\bm{y}_k^{t}|{\bm h}_k,t; \bm\Theta)$ denotes that the diffusion module receives ${\bm y}_k^t$ as input, conditioned on $t$ and $\bm h_k$. The denoising diffusion module is an MLP network composed of several residual blocks \cite{li2024autoregressiveimagegenerationvector}. Each block includes a LayerNorm, a linear layer, a SiLU activation, and a residual connection with an additional linear layer. Since the MLP denoising is conditioned on a vector $\bm{h}_k$ generated by the autoregressive language model, the rectified flow loss allows the update of the entire model parameters $\bm{\Theta}=\{\bm\Theta_{\text{LM}}, \bm\Theta_{\text{RF}}\}$ within a single-stage framework. To further accelerate diffusion convergence by enhancing supervision signals, directional information (e.g., cosine similarity between predicted and ground-truth vector fields) is incorporated \cite{lengREPAEUnlockingVAE2025,yaoReconstructionVsGeneration2025,yuRepresentationAlignmentGeneration2025}. Consequently, the auxiliary velocity direction loss is defined as
\begin{equation}
{\cal L}_{D}(\bm y_k,\bm h_k)  = 1 -\mathbb{E}_{t,{\bm y}_k^{0}} \left[ \cos{({\bm u}_k, v(\bm{y}_k^{t}|{\bm h}_k,t; \bm\Theta))}\right],
\end{equation}
where $\cos(\cdot, \cdot)$ denotes the cosine similarity operation. The timestep $t$ is commonly sampled from a uniform distribution ${\cal U}[0, 1]$. However, predicting the target vector filed ${\bm u}_k$ is more challenging for timesteps in the middle of $[0, 1]$, as the optimal prediction is simpler at the boundaries. To address this, we assign more weight to intermediate timesteps by sampling $t$ more frequently \cite{esser2024scalingrectifiedflowtransformers}, given as
\begin{equation}
\pi_{\text{ln}}(t; m, s) = \frac{1}{s \sqrt{2\pi} t (1 - t)} \exp \left( -\frac{(\text{logit}(t) - m)^2}{2s^2} \right),
\end{equation}
where $\text{logit}(t) = \log\frac{t}{1-t}$, with location parameter $m$ and scale parameter $s$. In this work, $m$ and $s$ are set to be 0 and 1 respectively.

\subsection{LM Guidance for Continuous-valued Token Generation}
Classifier-free guidance (CFG) \cite{ho2022classifierfreediffusionguidance} is widely used to improve the conditioning adherence of generative models. In diffusion models, the conditional and unconditional models share the same set of parameters and are jointly trained by randomly omitting the conditioning signal during training. During inference, the outputs of the conditional and unconditional models are combined using a hyperparameter, which controls the balance between diversity and fidelity in the synthesized speech. 

In the CLEAR framework, we implement CFG by jointly training the autoregressive LM and the MLP-based rectified flow head using both conditional and unconditional objectives. To train with CFG, we set all text embeddings to $ \mathbf{0} $ with a probability of 20\% in a mini-batch\footnote{An ablation study on the CFG strategy is conducted. Specifically, we found that setting text embeddings to $ \mathbf{0} $ at the LM prefix stage outperforms other CFG approaches,
such as randomly replacing the conditioning vector ${\bm h}_k$ with either \textbf{a)} a null embedding or \textbf{b)} a learnable embedding resulted in inferior performance.}, allowing the model to learn dual vector fields. At inference time, guided vector fields are derived through linear blending
\begin{equation}
\hat{v}(\bm{y}_k^{t}|{\bm h}_k,\bar{{\bm h}}_k,t; \bm\Theta) = w v(\bm{y}_k^{t}|{\bm h}_k,t; \bm\Theta) + (1-w)v(\bm{y}_k^{t}|\bar{{\bm h}}_k,t; \bm\Theta),
\end{equation}
where $\bar{{\bm h}}_k = f(\bm{y}_{<k},\bm{0})$ represents the conditioning vector calculated with $\textbf{0}$ text embedding, and $w$ is the guidance scale. 

\subsection{Variational Autoencoder for Continuous Audio Representation}\label{sec3:vae}
In the CLEAR framework, we present a deep compression variational autoencoder (VAE) to map the audio waveform to the continuous latent representation for efficient speech synthesis. The VAE architecture consists of a convolutional encoder and a convolutional causal decoder networks, enhanced with oobleck blocks \cite{evans2024fasttimingconditionedlatentaudio} and snake activation functions \cite{ziyinNeuralNetworksFail2020}. Given a 16 kHz audio waveform ${\bm a} \in \mathbb{R}^{L}$, where $L$ represents the total length of the audio, the VAE encoder compresses the waveform with a ratio of $D$ to produce a sequence of continuous audio representations, expressed as ${\bm y} = (\bm{y}_1, \bm{y}_2, \cdots, \bm{y}_{\lceil L / D \rceil})$. To ensure robust signal fidelity under aggressive downsampling of 16 kHz audio, parameter-free deep residual connections are utilized to address the optimization difficulty caused by high compression. These residual connections introduce non-parametric shortcuts \cite{chenDeepCompressionAutoencoder2025} into the autoencoder, enabling the network to focus on learning residuals through a space-to-channel transformation \cite{shi2016real}. Specifically, in the encoder's downsample blocks and decoder's upsample blocks, the non-parametric shortcut consists of a space-to-channel operation followed by a non-parametric channel averaging step to align the number of channels. Further details on shortcut connections can be found in the Appendix~\ref{apdix_shortcut} and  Appendix~\ref{apd:convsdis}. To facilitate stable training, a multi-task learning criterion with a discriminator setup \cite{defossezHighFidelityNeural2022} is employed and can be expressed as ${\cal L}_{\text{AE}} =  {\cal L}_{\text{ms}} + \alpha\cdot {\cal L}_{\text{fm}} + \beta \cdot {\cal L}_{\text{adv}} + \gamma \cdot {\cal L}_{\text{kl}}$,

where ${\cal L}_{\text{ms}}$ is the multi-resolution STFT loss \cite{steinmetzAutomaticMultitrackMixing2021}, ${\cal L}_{\text{fm}}$ is the feature matching loss from a multi-scale STFT discriminator \cite{defossezHighFidelityNeural2022}, ${\cal L}_{\text{adv}}$ is the adversarial loss \cite{goodfellowGenerativeAdversarialNetworks2014}, and ${\cal L}_{\text{kl}}$ is the 
Kullback-Leibler divergence loss. In this work, $\alpha$, $\beta$, and $\gamma$ are set to be $1.0$, $5.0$ and $0.1$, respectively. The downsampling ratio $D$ is set to $2048$ in this work, effectively balancing compression efficiency and reconstruction quality. Ablation studies can be found at Appendix~\ref{apd:compressratio}.

\subsection{Zero-shot CLEAR Text-to-Speech Synthesis}
\textbf{Single-stage Training Framework.} The training process of CLEAR is straightforward yet highly efficient. First, the text sequence $\bm x$ is converted into phonemes and mapped to text embeddings through an embedding layer. Meanwhile, speech audio $\bm a$ is compressed into continuous audio representations $\bm y$ using a VAE encoder. The input to the language model is then structured as $[S, \bm x, T, \bm y]$. A single end-to-end autoregressive model is trained using teacher forcing manner, optimizing the following loss: ${\cal L}={\cal L}_{\text{RF}}+{\cal L}_{\text{D}}+{\cal L}_{\text{s}}$\footnote{Since CLEAR produces continuous audio representations instead of discrete tokens, it cannot generate a stop token (e.g., <EOS>) to signal the end of generation for the autoregressive language model. Following the approach used in TransformerTTS \cite{liNeuralSpeechSynthesis2019} and SpeechT5 \cite{aoSpeechT5UnifiedModalEncoderDecoder2022}, a binary classifier with a fully connected layer is added to the language model's output to compute the binary cross-entropy loss, ${\cal L}_{s}$, for stop prediction.}. 

\textbf{In-context Learning for Zero-shot TTS.} 
Given the continuous representation of the audio prompt $\bm{\hat{y}}$ and its corresponding text prompt $\bm{\hat{x}}$, zero-shot speech synthesis is performed by autoregressively predicting the continuous audio representations $\bm{y}$ using the input sequence $[S, \bm{\hat{x}}, \bm{x}, T, \bm{\hat{y}}]$. In non-streaming speech synthesis, the continuous audio representations $\bm{y}$ are converted into speech audio via the VAE decoder once the autoregressive generation process is finished. In streaming speech synthesis, which reduces first-frame latency, the causal VAE decoder begins synthesizing speech immediately upon receiving the first chunk of continuous audio representations.

\vspace{-0.2cm}
\section{Experimental Setup}
\subsection{Training and Evaluation Dataset}
We utilize the open-source LibriHeavy \cite{kang2024libriheavy50000hoursasr} dataset to train the CLEAR model. LibriHeavy contains approximately 50,000 hours of speech data from 6,736 speakers and is derived from English audiobooks in the LibriVox project. In addition, the open-source LibriTTS \cite{zenLibriTTSCorpusDerived2019} dataset with 585 hours of speech data from 2456 distinct speakers is utilized to train the enhanced variational autoencoder in Sec.~\ref{sec3:vae}. For a fair comparison with state-of-the-art TTS systems, we adopt two established subsets from the open-source LibriSpeech(PC) test-clean set \cite{meisterLibriSpeechPCBenchmarkEvaluation2023,panayotovLibrispeechASRCorpus2015} for zero-shot TTS evaluation: \textbf{a)} Subset A from NaturalSpeech3 \cite{juNaturalSpeech3ZeroShot2024}, comprising 40 three-second audio prompts and 40 target samples, and \textbf{b)} Subset B from F5-TTS \cite{chen2024f5ttsfairytalerfakesfluent}, consisting of 40 audio prompts and 1,127 target samples.

\subsection{Experimental Settings}\label{sec42:config}

\textbf{Model Configurations:} The CLEAR-Base model consists of 24 Transformer blocks, each with 16 attention heads, an embedding dimension of 1024, and a feed-forward layer dimension of 4096. 
The MLP-based rectified flow head consists of 6 ResBlock1D layers, with 1024 dimensions across all layers. For the CLEAR-Large model, we scale up the embedding dimension to 1280 and the feed-forward layer dimension to 5120 for each layer. 


\textbf{Implementation Details:}
The CLEAR-Base model is trained with a batch size of 768 audio latents per GPU for 118 hours using 6 NVIDIA RTX 4090 24GB GPUs, with gradient accumulation set to 4. For the CLEAR-Large model, training is conducted with a batch size of 2048 audio latents per GPU for 112 hours on 2 NVIDIA H20 96GB GPUs. 
The models are optimized via the AdamW optimizer with momentum parameters $\beta_{1}=0.9$ and $\beta_{2}=0.95$. 
The learning rate follows a linear warm-up schedule, increasing to $10^{-4}$ over the first 1,000 updates, and remains constant thereafter.
The gradient norm clipping is set to 1. 
To expedite the training process, audio waveforms are resampled to 16 kHz and cached as latent representations before training. 
We utilize G2P to convert text transcripts into phonemes while preserving punctuation, which results in a phoneme vocabulary size of 203, with various punctuation patterns observed in the LibriHeavy dataset.
During training and inference, we apply exponential moving averaging (EMA) \cite{karrasAnalyzingImprovingTraining2024} on model weights. For inference, unless otherwise specified, we use a classifier-free guidance (CFG) scale of 2.5 and 10 denoising function evaluations. The details of the VAE training are provided in Appendix~\ref{apd:trainingVAE}.

\begin{table}[htbp]
    \centering
    \label{tab:sota}
    \caption{Objective evaluation results on the LibriSpeech (PC) test-clean Subset-A and Subset-B. ${\text{\ding{168}}}$ denotes the score reported in the corresponding baseline papers. ${\text{\ding{170}}}$ denotes the score reported in NaturalSpeech3 \cite{juNaturalSpeech3ZeroShot2024}. ${\text{\ding{169}}}$ denotes the score reported in F5-TTS \cite{chen2024f5ttsfairytalerfakesfluent}. ${\text{\ding{171}}}$ denotes the reproduced score. "Disc." and "Cont." denote "discrete-valued" and "continuous-valued" respectively. $\downarrow$ and $\uparrow$ represent that lower or higher values are better. }
    \resizebox{\textwidth}{!}{
    \begin{tabular}{lcccccc}
        \toprule
         \textbf{Model} & \textbf{Type} & \textbf{\#Params} & \textbf{Training Data} & \textbf{WER(\%)} $\downarrow$ & \textbf{SIM-o}$\uparrow$ & \textbf{UTMOS}$\uparrow$ \\
        \bottomrule
        \bottomrule
        \multicolumn{7}{c}{LibriSpeech test-clean \textbf{Subset} 
(Closed-sourced in \cite{chen2024valle2neuralcodec})}  \\ 
        \midrule
        Ground Truth$^{\text{\ding{168}}}$ & - & - & - & 2.20 & 0.75 & 4.09 \\ 
        \hdashline
        VALL-E 2\cite{chen2024valle2neuralcodec}$^{\text{\ding{168}}}$ & Disc. AR + Disc. NAR  & - & Libri-50k & 2.44 & 0.64 & - \\
        MELLE\cite{meng2024autoregressivespeechsynthesisvector}$^{\text{\ding{168}}}$ & Cont. AR & - & Libri-50k & 2.10 & 0.63 & - \\
        Voicebox\cite{le2023voicebox}$^{\text{\ding{168}}}$ & Cont. NAR & 330M & Libri-60k & 1.90 & 0.66 & -  \\ 
        \bottomrule
        \bottomrule
        \multicolumn{7}{c}{LibriSpeech test-clean \textbf{Subset-A} (Open-sourced in \cite{juNaturalSpeech3ZeroShot2024})}  \\ 
        \midrule
        Ground Truth$^{\text{\ding{171}}}$ & - & - & - & 1.94 & 0.68 & 4.09  \\ 
        VAE Reconstruction & - & - & - & 2.57 & 0.65 & 3.94 \\
        \hdashline
        VALL-E\cite{wang2023neural}$^{\text{\ding{170}}}$ & Disc. AR + Disc. NAR & 400M & Libri-60k & 6.11 & 0.47 & 3.68 \\
        MegaTTS\cite{jiangMegaTTS2Boosting2024}$^{\text{\ding{170}}}$ & Disc. AR + Cont. AR & 500M & Libri-60k & 2.32 & 0.53 & 4.02 \\
        NaturalSpeech3\cite{juNaturalSpeech3ZeroShot2024}$^{\text{\ding{170}}}$ & Disc. NAR & 500M & Libri-60k & 1.81 & \textbf{0.67} & \textbf{4.30} \\
        NaturalSpeech2\cite{shenNaturalSpeech2Latent2023}$^{\text{\ding{170}}}$ & Cont. NAR & 400M & Libri-60k & 1.94 & 0.55 & 3.88 \\
        DiTAR\cite{jia2025ditardiffusiontransformerautoregressive}$^{\text{\ding{170}}}$ & Cont. AR & 600M & Libri-60k & 1.78 & 0.64 & 4.15\\
        \hdashline
        \textbf{CLEAR-Base (ours)} & Cont. AR & 439M & Libri-50k & 1.83 & 0.55 & 4.21 \\
        \textbf{CLEAR-Large (ours)} & Cont. AR & 686M & Libri-50k & \textbf{1.74} & 0.56 & 4.26 \\
        \bottomrule
        \bottomrule
        \multicolumn{7}{c}{LibriSpeech-PC test-clean \textbf{Subset-B} (Open-sourced in \cite{chen2024f5ttsfairytalerfakesfluent})}  \\ 
        \bottomrule
        Ground Truth$^{\text{\ding{171}}}$ & - & - & - & 2.47 & 0.69 & 4.09 \\ 
        VAE Reconstruction & - & - & - & 2.89 & 0.63 & 4.08 \\ 
        \hdashline
        CosyVoice\cite{du2024cosyvoicescalablemultilingualzeroshot}$^{\text{\ding{169}}}$ & Disc. AR + Cont. NAR & 300M & Multi-170k & 3.59 & 0.66 & - \\
        CosyVoice 2\cite{du2024cosyvoice2scalablestreaming} & Disc. AR + Cont. NAR & 500M & Multi-170k & 2.47 & 0.65 & \textbf{4.35} \\
        FireRedTTS\cite{guo2024fireredtts}$^{\text{\ding{169}}}$ & Disc. AR + Cont. NAR & 580M & Multi-248k & 2.69 & 0.47 & - \\  
        MaskGCT\cite{wangMaskGCTZeroShotTexttoSpeech2024a}$^{\text{\ding{168}}}$ & Disc. NAR & 1048M & Emilia-100k & 2.72 & \textbf{0.69} & 3.90 \\
        F5-TTS\cite{chen2024f5ttsfairytalerfakesfluent}$^{\text{\ding{168}}}$ & Cont. NAR & 300M & Emilia-100k & 2.42 & 0.66 & 3.88 \\ 
        DiTAR\cite{jia2025ditardiffusiontransformerautoregressive}$^{\text{\ding{168}}}$ & Cont. AR & 600M & Emilia-100k & 2.39 & 0.67 & 4.22\\ 
        \hdashline
        \textbf{CLEAR-Base (ours)} & Cont. AR & 439M & Libri-50k & 2.21 & 0.59 & 4.22 \\
        \textbf{CLEAR-Large (ours)} & Cont. AR & 686M & Libri-50k & \textbf{1.88} & 0.59 & 4.22 \\
        \bottomrule
     \end{tabular}}
     \vspace{-0.6cm}
\end{table}
\subsection{Evaluation Metrics}
\label{section:evaluation_metrics}
The proposed CLEAR model is evaluated on a cross-sentence TTS task using both objective and subjective metrics, including robustness, speaker similarity, naturalness, and efficiency.

\textbf{Objective Evaluation.} \textbf{a) Word error rate} (WER) used to assess robustness. For consistency, following the configurations in \cite{chen2024f5ttsfairytalerfakesfluent,juNaturalSpeech3ZeroShot2024}, we perform speech recognition on the synthesized speech via Hubert-based model and Faster-whisper-large-v3 \cite{radford2023robust} for the subset A of LibriSpeech test-clean and subset B of LibriSpeech-PC test-clean, respectively;
\textbf{b) Speaker similarity} between the generated and the original target speeches (SIM-o). Specifically, the WavLM-TDNN \cite{chen2022wavlm} is leveraged to extract corresponding speaker features to compute their cosine similarity; \textbf{c) UTMOS} \cite{saeki2022utmos} that is used to predict the mean opinion score (MOS) to assess the speech naturalness;
and \textbf{d) Real-time factor} (RTF) used to evaluate the inference efficiency. We measure the average inference time through generating one second speech segments for all synthesized samples.

\textbf{Subjective Evaluation}. Four types of mean opinion score (MOS) metrics are adopted, including \textbf{a) N-MOS} to assess speech naturalness, \textbf{b) Q-MOS} for perceived sound quality, \textbf{c) S-MOS} to measure speaker voice similarity, and \textbf{d) CMOS} for side-by-side comparison with human speech.

\vspace{-0.2cm}
\section{Experimental Results}
\subsection{Objective Evaluation of Zero-shot Speech Synthesis}

Table~1 presents an objective performance comparison between the proposed CLEAR model and recent state-of-the-art TTS systems, several trends can be observed.

\textbf{a)} On the LibriSpeech Subset-A evaluation set of NaturalSpeech3 \cite{juNaturalSpeech3ZeroShot2024}, \textbf{1)} the proposed CLEAR-Base model shows significant improvements over conventional discrete-valued AR-based systems, such as VALL-E and MegaTTS, with comparable model parameters. Specifically, it delivers up to \textbf{4.28\%} absolute WER reductions, improves speaker similarity scores by up to \textbf{0.08}, and achieves an impressive UTMOS score of \textbf{4.21}. This demonstrates that the proposed continuous VAE latents, which bypass manually designed strategies for discrete codec codes, produce more stable and natural speech. \textbf{2)} When compared to continuous-valued NAR NaturalSpeech2 system, the CLEAR-Base model achieves better performance in WER and UTMOS, while maintaining comparable speaker similarity. In addition, it matches the WER and UTMOS performance of the discrete-valued NAR NaturalSpeech3 system but lags in speaker similarity. This difference may arise from the rich speaker information embedded in the discrete value tokens extracted by FACodec in NaturalSpeech3, which incorporates a speaker classification task by predicting speaker IDs.

\textbf{b)} On the LibriSpeech Subset-B evaluation set of F5-TTS \cite{chen2024f5ttsfairytalerfakesfluent}, the CLEAR-Base model outperforms both the two-stage cascaded TTS models (CosyVoice, CosyVoice-2, and FireRedTTS) and the state-of-the-art NAR F5-TTS systems. Specifically, it achieves the lowest WER and an excellent UTMOS score. Although a decline in speaker similarity is observed, the CLEAR-Base model delivers competitive subjective speaker voice similarity (S-MOS), as detailed in Section \ref{section:subjective}. Additional analysis of the speaker similarity metrics is provided in Appendix~\ref{apd:sim}.

\textbf{c)} Compared to the state-of-the-art continuous-valued DiTAR model, the proposed CLEAR-Large model demonstrates enhanced robustness and audio quality in both the LibriSpeech Subset-A and Subset-B evaluation sets. Specifically, the CLEAR-Large model achieves absolute WER reduction of \textbf{0.51\%} (\textbf{21.3\%} relative) over DiTAR, while maintaining high audio quality with a UTMOS score of \textbf{4.22} on the Subset-B dataset.

\begin{table}[htbp]
    \centering
    \label{tab:efficiency}
    \vspace{-0.2cm}
    \caption{Average AR decoding steps (Avg. Decoding Steps) and real-time factor (RTF) through generating a 10-second speech segment. ${\text{\ding{168}}}$ denote the score quoted from VALL-E R \cite{han2024vall}. ${\text{\ding{170}}}$ denote the score quoted from CLaM-TTS \cite{kim2024clam}. ${\text{\ding{169}}}$ denote the score reported in the baseline paper.}
    \resizebox{0.7\textwidth}{!}{
    \begin{tabular}{lccc}
        \toprule
        \textbf{Model} & \textbf{Type} & \textbf{Avg. Decoding Steps} & \textbf{RTF}$\downarrow$ \\ \bottomrule
        VALL-E \cite{wang2023neural} $^{\text{\ding{168}}}$ & Disc. AR + Disc. NAR & 750 & 1.03 \\
        VALL-E 2 \cite{chen2024valle2neuralcodec} $^{\text{\ding{168}}}$ & Disc. AR + Disc. NAR & 750 & 0.73 \\
        CosyVoice \cite{du2024cosyvoicescalablemultilingualzeroshot} $^{\text{\ding{169}}}$ & Disc. AR + Cont. NAR & 250 & 0.92 \\
        VALL-E R \cite{han2024vall} $^{\text{\ding{168}}}$ & Disc. AR & 375 & 0.37 \\
        Voicebox \cite{le2023voicebox}$^{\text{\ding{170}}}$ & Cont. NAR & - & 0.64  \\
        CLaM-TTS \cite{kim2024clam} $^{\text{\ding{170}}}$ & Cont. NAR & - & 0.42 \\
        F5-TTS \cite{chen2024f5ttsfairytalerfakesfluent} $^{\text{\ding{169}}}$ & Cont. NAR & - & 0.31 \\
        DiTAR \cite{jia2025ditardiffusiontransformerautoregressive} $^{\text{\ding{169}}}$ & Cont. AR & 400 & - \\
        MELLE \cite{meng2024autoregressivespeechsynthesisvector}$^{\text{\ding{169}}}$ & Cont. AR & 620 & 0.55 \\
        \hdashline
        \textbf{CLEAR-Base (ours)} & Cont. AR & \textbf{78} & \textbf{0.18} \\ 
        \textbf{CLEAR-Large (ours)} & Cont. AR & \textbf{78} & \textbf{0.29} \\ 
        \bottomrule
    \end{tabular}}
    \vspace{-0.5cm}
\end{table}
Further comparison of inference efficiency against recent TTS systems is presented in Table~2. The CLEAR-Base model achieves the best RTF performance among all TTS systems, including both AR and NAR frameworks. Even with increased model parameters, the proposed CLEAR-Large model maintains superior inference efficiency, making it ideal for real-time applications. The low-latency capability of our model is achieved through two key innovations. First, the enhanced VAE generates more compact continuous latents compared to 
the lengthy discrete-valued tokens (bitrate constraints) and continuous mel-spectrograms features. This reduces the required autoregressive decoding steps to as low as 7.8, leading to significantly faster inference. Second, a simplified MLP-based rectified flow head is introduced. By assigning most parameters to the language model, it keeps the diffusion model lightweight and efficient while preserving high-quality synthesis. Further analysis of training efficiency is available in Appendix~\ref{apd:trainingEffi}.

\begin{table}[htbp]
    \centering
    \label{tab:subjective}
    \caption{Subjective evaluation results on LibriSpeech-PC test-clean Subset-B \cite{chen2024f5ttsfairytalerfakesfluent}. }
    \resizebox{0.8\textwidth}{!}{
    \begin{tabular}{lccccc}
        \toprule
        \textbf{Model} & \textbf{Type} & \textbf{N-MOS} &  \textbf{Q-MOS} & \textbf{S-MOS} & \textbf{CMOS} \\ 
        \bottomrule
        Ground Truth & - & 4.01 & 4.00 & 3.68 & 0.00 \\
        \hdashline
        F5-TTS \cite{chen2024f5ttsfairytalerfakesfluent} & Cont. NAR & 3.75 & 3.45 & \textbf{4.03} & -0.17 \\
        CosyVoice-2 \cite{du2024cosyvoice2scalablestreaming} & Disc. AR + Cont. NAR & 3.90 & 4.10 & 3.76 & \textbf{+0.25} \\
        \hdashline
        \textbf{CLEAR-Base (ours)} & Cont. AR & \textbf{4.09} & \textbf{4.14} & 4.02 & +0.04 \\
        \bottomrule
    \end{tabular}}
\end{table}

\subsection{Subjective Evaluation of Zero-shot Speech Synthesis}
\label{section:subjective}
Subjective evaluation metrics of Section \ref{section:evaluation_metrics} including N-MOS, Q-MOS, S-MOS and CMOS are assessed based on a human rating system. More details are provided in Appendix~\ref{apd:subject}. 
Two open-sourced state-of-the-art TTS models, including the cascaded CosyVoice-2 system and the NAR F5-TTS system, are adopted to generate audio samples in the same configuration for a fair comparison.
 
As presented in Table~3, the proposed CLEAR-Base model achieves best performance in N-MOS (naturalness) and Q-MOS (quality) compared to CosyVoice2, F5-TTS, and even ground truth audio, 
indicating its capability to generate natural, high-quality speech rivaling human performance. 
Furthermore, the CLEAR model demonstrates speaker similarity performance comparable to F5-TTS, while outperforming CosyVoice-2 and even ground truth audio, showcasing its in-context learning capability to mimic a speaker's voice based on an audio prompt.

\begin{table}[t!]
    \centering
    \label{tab:streaming}
    \caption{Performance of the proposed CLEAR-Base model evaluated on LibriSpeech Subset-B when running in a streaming synthesis manner. "FFL" denotes first-frame latency for speech synthesis.}
    \resizebox{0.7\textwidth}{!}{
    \begin{tabular}{cc|cccc}
        \hline
        \textbf{Chunk Size $\bm{\Omega}$} & \textbf{Overlap $\bm{\Psi}$} & \textbf{WER(\%)}$\downarrow$ & \textbf{SIM-o}$\uparrow$ & \textbf{UTMOS}$\uparrow$ & \textbf{FFL (ms)} $\downarrow$  \\
        \hline
        4 & 1 & 2.34 & 0.57 & 4.27 & 96 \\
        8 & 2 & 2.24 & 0.57 & 4.27 & 191 \\
        16 & 4 & 2.18 & 0.58 & 4.27 & 383 \\
        32 & 8 & 2.19 & 0.59 & 4.26 & 766 \\
        \hdashline
        \multicolumn{2}{c|}{Non-streaming Synthesis} & 2.21 & 0.59 & 4.26 & -\\
        \hline
    \end{tabular}}
    \vspace{-0.3cm}
\end{table}
\vspace{-0.1cm}
\subsection{Streaming Speech Synthesis}
Streaming capability is vital for real-world applications (i.e., virtual assistants, audiobooks, and public announcements) as it drastically reduces latency and enhances the user experience. Users can begin hearing speech almost immediately, creating a more natural and conversational interaction, similar to human speech. The rectified flow head empowers our model to achieve low-latency streaming synthesis by allowing the diffusion process to function independently for each token, eliminating the need to generate the entire output beforehand in DiT. In the streaming speech synthesis process, the rectified flow head sequentially generates tokens, which are accumulated into chunks of a predefined size, denoted as $\bm{\Omega}$. Once the number of tokens in a chunk reaches this threshold, the chunk is passed to the causal VAE decoder to produce the corresponding waveform. To ensure smooth transitions, a fade-in-fade-out technique is applied to the overlapping region, denoted as $\bm{\Psi}$, of the waveform. The first-frame latency (FFL) refers to the time interval between the start of system processing and the moment when the first synthesized audio waveform becomes audible, which is utilized to measure the latency performance. Table~4 shows the performance of the proposed CLEAR-Base model when running in a streaming speech synthesis manner. With a chunk size of 4, the proposed CLEAR model achieves a mere 96ms latency on a 4096-GPU while matching non-streaming synthesis performance in robustness, speaker similarity, and quality. Specifically, its WER of 2.21\% and UTMOS of 4.27 in a streaming fashion, significantly outperform most non-streaming TTS systems shown in Table~1.

\vspace{-0.3cm}
\section{Conclusion}
This paper presents CLEAR, a novel zero-shot text-to-speech (TTS) framework that directly models continuous audio representations, addressing key limitations of traditional autoregressive (AR) TTS systems based on discrete audio tokens. By leveraging an enhanced variational autoencoder with shortcut connections, CLEAR achieves a high compression ratio, mapping waveforms into compact continuous latents, while a lightweight rectified flow head efficiently models the continuous latent probability distribution. Experimental results demonstrate that CLEAR delivers competitive performance in terms of robustness, speaker similarity, and naturalness while achieving a lower real-time factor compared to state-of-the-art TTS models. Moreover, CLEAR supports streaming speech synthesis with a first-frame delay as low as 96ms, without compromising speech quality.

\bibliographystyle{plainnat}
\bibliography{cite_jj}

\newpage
\appendix
\section{Limitations} \label{apd:limitations}
Although CLEAR achieves competitive results with an efficient architecture, we identify several areas for potential improvement. First, although CLEAR demonstrates promising subjective speaker similarity ratings, its objective speaker similarity metrics fall short compared to other state-of-the-art models, highlighting opportunities for refinement. Future work will focus on integrating speaker embeddings into both CLEAR's language model and rectified flow components to improve voice characteristic fidelity. Second, the current evaluation is limited to English, using the LibriSpeech test set. Future research will explore multilingual settings and more diverse datasets. Third, the existing CLEAR model generates autoregressive hidden states sequentially. Enhancing CLEAR to support multi-token prediction could significantly improve inference efficiency.

\section{Broader Impact and Ethics Statement} \label{broaderImpact}
Our research on CLEAR aims to advance the field of text-to-speech synthesis by introducing a more efficient and accessible approach to audio generation. 

First, by eliminating the need for discrete tokens and heavy codebooks, CLEAR significantly reduces computational requirements and energy consumption in both training and inference, contributing to more environmentally sustainable AI development. 

Second, the low-latency and streaming capabilities of CLEAR (96ms first-frame delay) make it particularly valuable for real-time applications, potentially benefiting assistive technologies for individuals with speech or reading disabilities, educational tools for language learning, real-time translation services, and healthcare applications requiring immediate voice feedback.

We acknowledge potential risks and challenges. Like other TTS systems, CLEAR's ability to mimic voices could be misused for impersonation or fraud. To mitigate this, we recommend implementing speaker consent protocols, developing robust voice authentication systems, and integrating speech synthesis detection mechanisms. More specifically, (a) Our model release will be accompanied by a clear usage policy and ethical guidelines prohibiting malicious applications; (b) We will implement a verification system requiring users to agree to terms of use before accessing the model; (c) The released model will include built-in watermarking for generated audio to ensure traceability; (d) We limit the length and content of generated speech through content filtering mechanisms; (e) Model weights will be released through a controlled academic repository rather than public hosting platforms.

\section{Variational Autoencoder Analysis}\label{apd:vae}
\subsection{Deep Shortcut Connection in Waveform Variational Autoencoder}\label{apdix_shortcut}
Inspired by the image compression in \cite{chenDeepCompressionAutoencoder2025}, the parameter-free shortcut connection is utilized in the waveform variational autoencoder (VAE). The enhanced VAE with shortcut connection shows high reconstruction quality even under a high compression ratio based on our experiments in Figure \ref{fig:quantization} and Table~1 (VAE reconstruction results). Specifically, the enhanced VAE with shortcut connections achieves a word error rate (WER) of 2.89\%, a SIM-o score of 0.63, and a UTMOS of 4.08, which are comparable to the ground truth audio scores of WER 2.47\%, SIM-o 0.69, and UTMOS 4.09 on the LibriSpeech-PC test-clean Subset-B. 

\begin{figure}[htbp]
    \centering
    \includegraphics[width=0.7\linewidth]{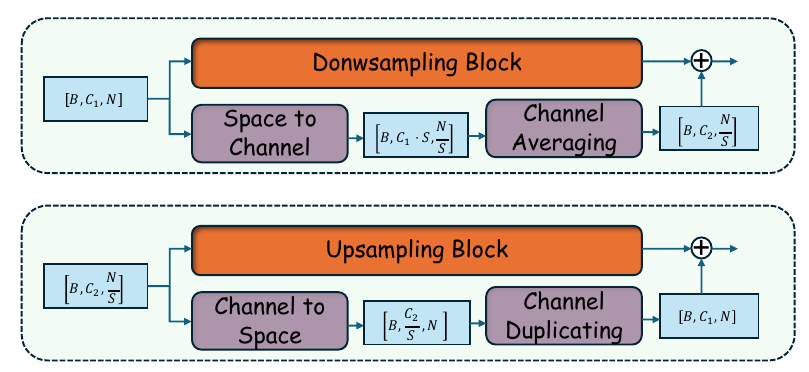}
    \caption{The parameter-free shortcut connection used in waveform VAEs. For space-to-channel and channel-to-space operations, the input is reshaped to match the time dimension of the output. Subsequently, averaging or duplication is applied to align the channel dimension with the output.}
    \label{fig:dc}
\end{figure}

Figure~\ref{fig:dc} illustrates the shortcut connection operations in our VAE. During each downsampling or upsampling operation in the VAE, reshaping is performed along with channel averaging or channel duplication to ensure the output shape matches that of the block.

\subsection{Implementation Details of Waveform VAE} \label{apd:trainingVAE}
The waveform VAE is trained with KL regularization and a downsampling ratio of 2048. The strides of the convolution layers in the encoder are set to $[2, 4, 4, 8, 8]$ and reversed for the decoder. The causal decoder architecture employs causal convolutions in place of standard convolution layers. For optimization, we use learning rates of $1.5 \times 10^{-4}$ and $3 \times 10^{-4}$ for the generator and discriminator, respectively, with a warmup phase of 25,000 steps before applying adversarial training. The weights of both the generator and discriminator are updated using the AdamW optimizer, with momentum parameters $\beta_1 = 0.8$ and $\beta_2 = 0.99$, and a weight decay of 0.001. We use AutoClip\cite{seetharamanAutoClipAdaptiveGradient2020} to perform gradient clipping during the training for VAEs.

\subsection{Continuous-valued and Discrete-valued VAEs Convergence Under High Compression} \label{apd:convsdis}
\begin{table}[htbp]
    \centering
    \caption{Quantization configurations for different VQ-VAE models. Only the codebook size is adjusted to ensure a fair comparison between VQ and FSQ.}
    \begin{tabular}{ccccc}
        \toprule
         \textbf{Quantization} & \textbf{Codebook Size} & \textbf{Codebook Dim.} & \textbf{Levels} & \textbf{Num. of Quantizers} \\
         \bottomrule
         VQ & 1024 & 128 & - & 1 \\
         FSQ & 4375 & 128 & $[7, 5, 5, 5, 5]$ & 1 \\
         RVQ & 1024 & 128 & - & 8 \\
         RFSQ & 4375 & 128 & $[7, 5, 5, 5,5]$ & 8 \\
         \bottomrule
    \end{tabular}

    \label{tab:vaes}
\end{table}
We have trained i) continuous-valued VAE with shortcut connection using a KL regularization, and ii) discrete-valued VAE using a discrete quantization-based regularization, including Vector Quantization(VQ)\cite{vandenoordNeuralDiscreteRepresentation2017}, Residual Vector Quantization(RVQ)\cite{kumarHighFidelityAudioCompression2023}, Finite Scalar Quantization(FSQ)\cite{mentzerFiniteScalarQuantization2023} and Residual Finite Scalar Quantization(RFSQ), under high-compression ratio of 4096 scenario. Table~\ref{tab:vaes} shows the detailed configuration for different quantization methods in our experiments. All VAEs share identical model configurations and training strategies described in the Appendix~\ref{apd:trainingVAE}. For learnable codebooks, we use a commitment loss weight of 0.25 and a codebook loss of 1.0. 

\begin{figure}[htbp]
    \centering
    \includegraphics[width=0.7\linewidth]{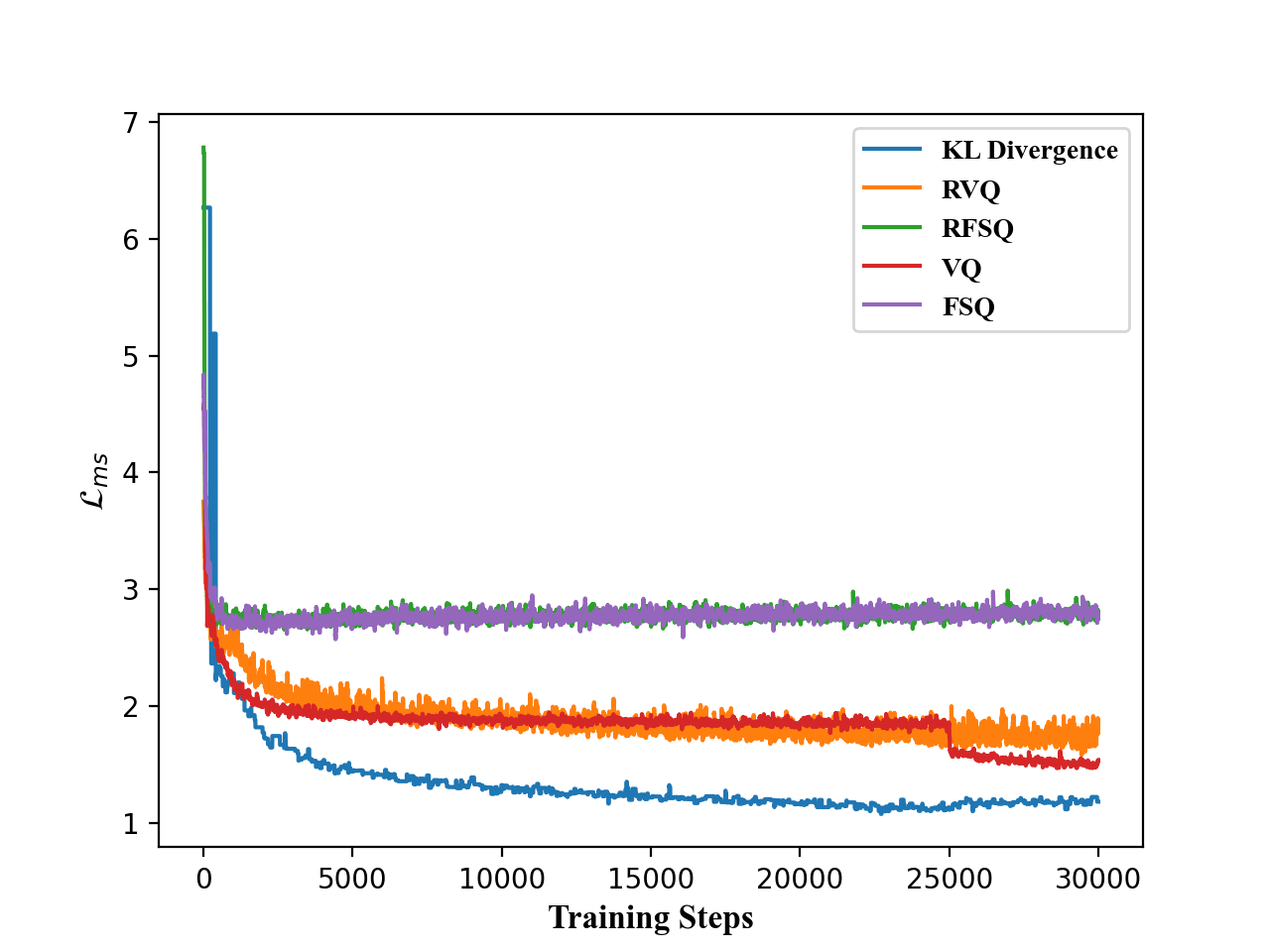}
    \caption{Model convergence of our continuous VAE (KL divergence regularization) and discrete VAEs (RVQ, RFSQ, VQ, FSQ) under a high compression ratio of 4096.}
    \label{fig:quantization}
\end{figure}
Figure \ref{fig:quantization} compares the training convergence of the continuous-valued VAE with shortcut connection used in the proposed CLEAR model with discrete-valued VAEs commonly used in conventional autoregressive (AR) based text-to-speech (TTS) synthesis systems, under a high compression ratio of 4096. As illustrated in this figure, the proposed continuous-valued VAE demonstrates faster and better convergence in terms of the multi-resolution STFT loss compared to other discrete-valued approaches. 

\subsection{Zero-shot TTS Performance of CLEAR with VAEs Under Different Compression Ratio}
\subsubsection{Model Configuration}
In this work, we present three configurations for CLEAR, leading to CLEAR-Small, CLEAR-Base, and CLEAR-Large. The detailed hyperparameters are provided in Table~\ref{tab:hyper}.
\begin{table}[htbp]
    \centering
    \caption{Configurations of CLEAR of different sizes.}
    \begin{tabular}{llccc}
        \toprule
         \multicolumn{2}{c}{\textbf{Model Config}} & \textbf{CLEAR-Small} & \textbf{CLEAR-Base} & \textbf{CLEAR-Large} \\
         \bottomrule
         \multirow{5}{*}{\textbf{Language Model}} & \# Param. & 227M & 403M & 630M \\
         & Number of layers & 24 & 24 & 24 \\
         & Hidden dim & 768 & 1024 & 1280 \\
         & Number of heads & 12 & 16 & 20 \\
         & FFN dim & 3072 & 4096 & 5120 \\
         \bottomrule
         \multirow{3}{*}{\textbf{Rectified Flow Head}} & \# Param. & 20M & 36M & 56M \\
         & Number of layers & 6 & 6 & 6 \\
         & Hidden dim & 512 & 1024 & 1280 \\
         \bottomrule
    \end{tabular}
    \label{tab:hyper}
\end{table}

\subsubsection{Analysis of Different Compression Ratio}\label{apd:compressratio}
To examine the effect of continuous latents with different compression ratios on zero-shot TTS, we first extracted continuous audio latents from VAEs configured with downsampling ratios of 768, 2048, and 4096\footnote{The convolution strides for downsampling ratios of 768, 2048, and 4096 are $[2,4,4,4,6]$, $[2,4,4,8,8]$, and $[2,2,4,4,8,8]$, respectively.}. Using these latents, we trained the CLEAR-Small model on the 960-hour LibriSpeech dataset. The training and inference procedures followed the same configuration as described in Section~\ref{sec42:config}.

Table~\ref{tab:clear_diff_down} shows the performance of CLEAR-Small model using continuous audio latents with different downsampling ratios. As compression ratios increase, we observe a trade-off between speech robustness and speaker similarity. Higher compression ratios lead to shorter continuous latent sequences, which while beneficial for LM semantic modeling, result in decreased WER. Conversely, lower compression ratios produce longer sequences that, while potentially suffering from error accumulation in AR modeling causing pronunciation issues, better preserve acoustic features including speaker characteristics, thus improving speaker similarity scores. To strike a balance between speech robustness and speaker similarity, we employ a downsampling ratio of 2048 in this paper.

With a downsampling rate of 2048, our VAE effectively balances quality and efficiency. For instance, in 16kHz speech, a mel spectrogram with a hop size of 160 generates 100 latents per second, but the VAE only requires $16000/2048 \approx 7.8$ latents per second. This drastically reduces the number of AR decoding steps, resulting in a significant boost in inference efficiency.

\begin{table}[htbp]
    \centering
    \caption{Performance of CLEAR-Small with autoencoders at different downsampling ratios.}
    \begin{tabular}{cccc}
        \toprule
        \textbf{Model} & \textbf{Downsampling Ratio} & \textbf{WER}$\downarrow$ & \textbf{SIM-o}$\uparrow$ \\
        \bottomrule
        \multirow{3}{*}{CLEAR-Small} & 768 & 15.4 & \textbf{0.46} \\
        & 2048 & 5.74 & 0.44 \\
        & 4096 & \textbf{4.90} & 0.39 \\
        \bottomrule
    \end{tabular}
    \label{tab:clear_diff_down}
\end{table}

\section{Analysis of Zero-shot CLEAR TTS Model}\label{apd:ttsmodel}
\subsection{Training Efficiency} \label{apd:trainingEffi}

\begin{figure}[htbp]
  \centering
  \includegraphics[width=0.9\textwidth]{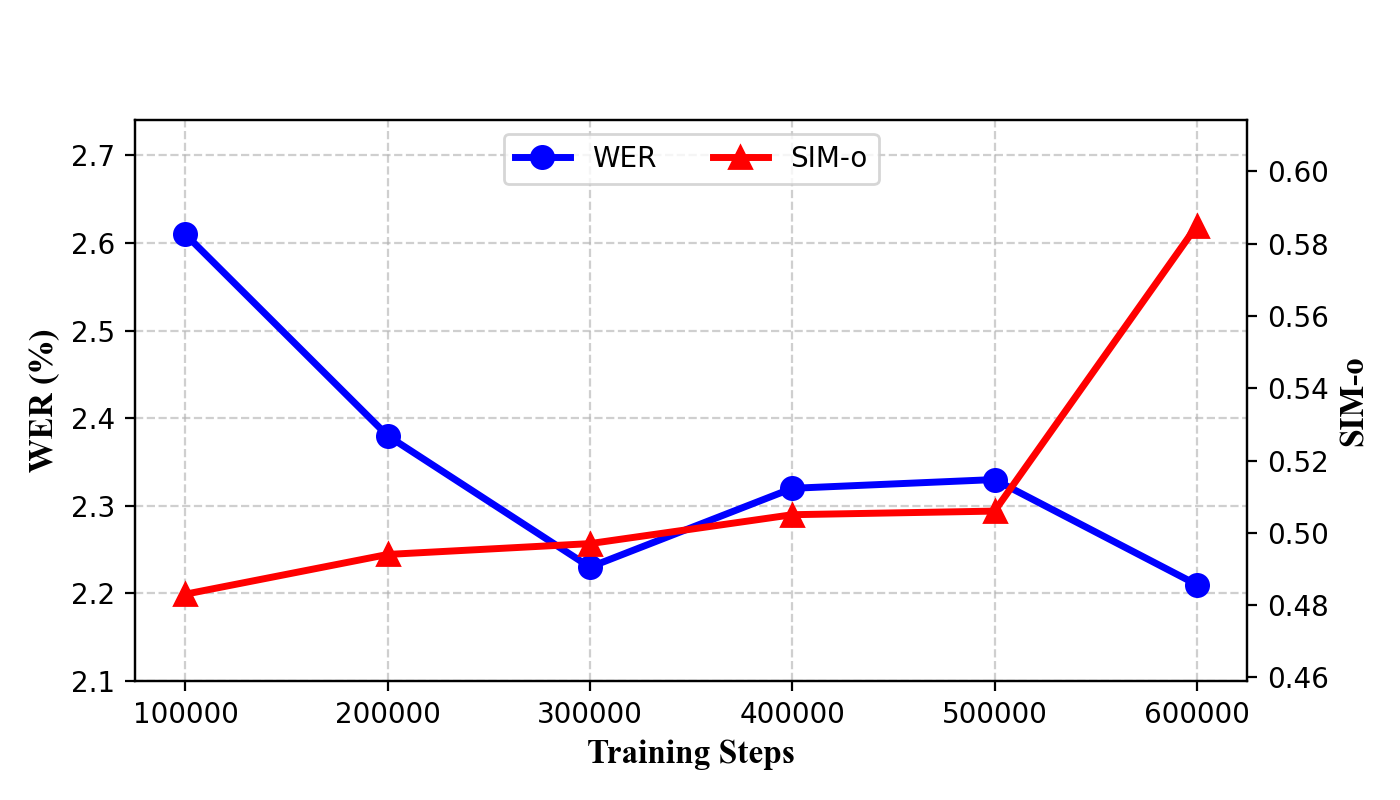}
  \caption{CLEAR-Base model performance (WER(\%)/SIM-o) versus training steps.}
  \label{fig:training_steps}
\end{figure}

Figure \ref{fig:training_steps} demonstrates the variation in CLEAR-Base's WER and SIM-o scores throughout the training process evaluated on the LibriSpeech-PC test-clean Subset-B. Both WER and SIM-o metrics exhibit consistent improvement as training progresses, achieving a good performance at 600k steps. It's worth noting that the highly compressed VAE latent representations in this work offer two significant advantages. First, the compact token sequences enable the language model to capture semantic information more efficiently, achieving rapid convergence in robustness (2.22\% WER in just 300k steps). Second, the shorter sequence length substantially reduces computational requirements, with our CLEAR-Base model achieving impressive results using only 24GB of GPU memory (completed 600k training steps with 50k-hrs training data in 118 hours using 6 NVIDIA RTX 4090 24GB GPUs), making it particularly suitable for resource-constrained scenarios.

\subsection{Analysis of Speaker Similarity Performance}\label{apd:sim}
In addition to the widely used WavLM-TDNN, we also evaluate the SIM score of CLEAR using WavLM-base-sv\footnote{https://huggingface.co/microsoft/wavlm-base-sv}. The results indicate that variations in the SIM score are impacted by biases in the speaker embedding model. When evaluated with WavLM-base-sv, The CLEAR-Base demonstrates a smaller SIM score gap compared to the SOTA open-sourced models. 

Despite our strong performance in subjective speaker similarity ratings in Table~\ref{tab:subjective}, there remains a gap between CLEAR and other SOTA models in objective speaker similarity metrics, which requires further improvement. We have identified two main limitations:

\textbf{a)} Regarding conditioning strategies in different latent generation baselines, most existing zero-shot TTS systems directly inject intermediate representations of the prompt audio (such as mel-spectrograms in F5-TTS \cite{chen2024f5ttsfairytalerfakesfluent}, discrete tokens in NaturalSpeech3 \cite{juNaturalSpeech3ZeroShot2024}, or speaker embeddings in CosyVoice \cite{du2024cosyvoicescalablemultilingualzeroshot}) into the generative network. CLEAR, however, the speaker characteristics must be inferred by the language model (LM) from the input audio prompt through in-context learning. This constrains the RF head's ability to generate accurate speech latents that match the prompt.

\textbf{b)} While the shorter continuous latent sequences resulting from high compression rates benefit autoregressive modeling shown in Table~\ref{tab:clear_diff_down}, they are less effective at preserving voice characteristics such as timbre compared to longer mel-spectrogram features.

Therefore, in future work, we plan to explore incorporating speaker embeddings into CLEAR to guide both the LM and the rectified flow head toward more faithful reproduction of prompt voice characteristics.

\begin{table}[htbp]
    \centering
     \caption{Speaker similarity results using two different speaker verification models on LibriSpeech test-clean Subset-B.}
    \begin{tabular}{ccc}
        \toprule
         \textbf{Model} & \textbf{WavLM-TDNN}\footnote{https://github.com/microsoft/UniSpeech} & \textbf{WavLM}\footnote{https://huggingface.co/microsoft/wavlm-base-sv} \\
         \bottomrule
         CosyVoice 2 \cite{du2024cosyvoice2scalablestreaming} & 0.65 & 0.956 \\
         F5-TTS \cite{chen2024f5ttsfairytalerfakesfluent} & 0.66 & 0.958 \\
         CLEAR-base & 0.59 & 0.950 \\
         \bottomrule
    \end{tabular}
    \label{tab:sim}
\end{table}

\begin{figure}[htbp]
    \centering
    \includegraphics[width=0.7\textwidth]{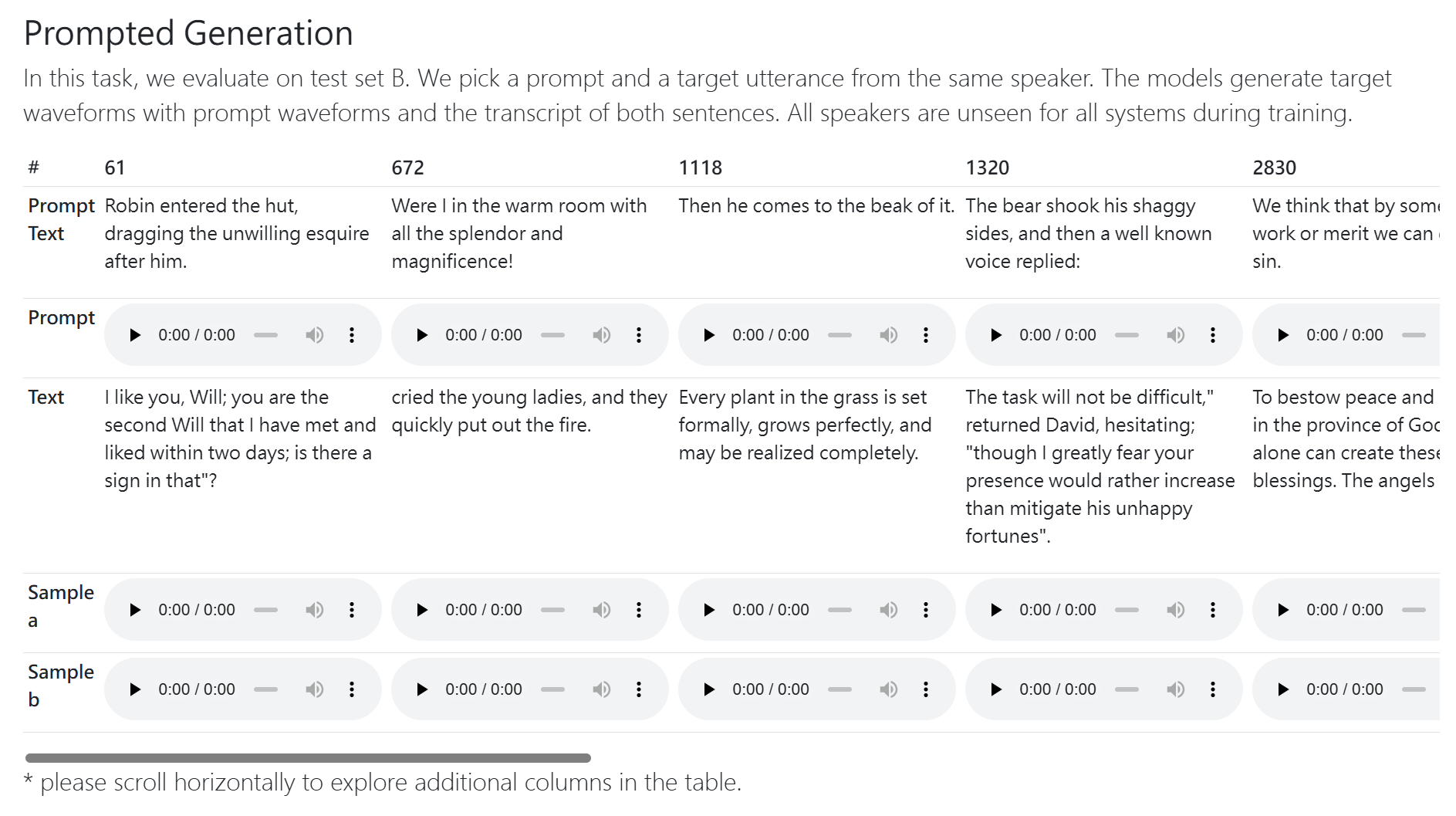}
    \caption{The user interface for subjective evaluation.}
    \label{fig:mos_ui}
\end{figure}
\begin{figure}[htbp]
    \centering
    \includegraphics[width=0.5\textwidth]{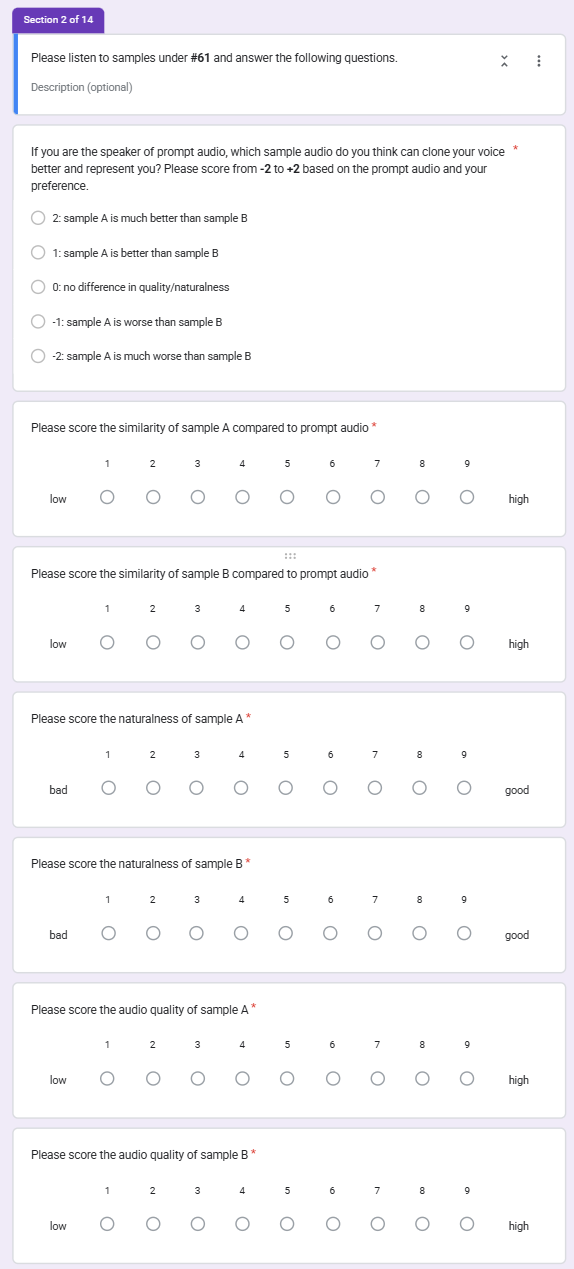}
    \caption{The questions of each samples pair for subjective evaluation. When calculating the MOS score, we rescale Q-MOS, S-MOS and N-MOS to have a maximum score of 5.}
    \label{fig:mos_q}
\end{figure}

\section{Subjective Evaluation}\label{apd:subject}
We supplement objective metrics with comprehensive subjective evaluations to assess the zero-shot TTS performance. The evaluation protocol requires human raters to compare two audio samples against a reference, providing both comparative scores (CMOS) between samples and individual ratings (N-MOS, Q-MOS, S-MOS) for each audio.

Specifically, the N-MOS (speech naturalness), Q-MOS (speech quality), S-MOS (speaker voice similarity) evaluate synthesized audio on a scale of 1 to 9, which is subsequently mapped to a 1-5 scale with 0.5 step intervals.
For CMOS (side-by-side human speech comparison), human evaluators compare the synthesized audio to the ground truth (GT) and assign a score ranging from -2 to 2. The evaluation interface and question format are shown in Figures~\ref{fig:mos_ui} and Figure~\ref{fig:mos_q}, respectively.

\end{document}